# A Cascaded Iterative Fourier Transform Algorithm For Optical Security Applications

Guohai Situ, Jingjuan Zhang

Department of Physics, Graduate School of the Chinese Academy of Sciences, P. O. Box 3908, Beijing, 100039, China

**Abstract**: A cascaded iterative Fourier transform (CIFT) algorithm is presented for optical security applications. Two phase-masks are designed and located in the input and the Fourier domains of a 4-f correlator respectively, in order to implement the optical encryption or authenticity verification. Compared with previous methods, the proposed algorithm employs an improved searching strategy: modifying the phase-distributions of both masks synchronously as well as enlarging the searching space. Computer simulations show that the algorithm results in much faster convergence and better image quality for the recovered image. Each of these masks is assigned to different person. Therefore, the decrypted image can be obtained only when all these masks are under authorization. This key-assignment strategy may reduce the risk of being intruded.

## 1. Introduction

Optical techniques have shown great potential in the field of information security applications. Recently Réfrégier and Javidi [1] proposed a novel double-random-phase encoding technique, which encodes a primary image into a stationary white noise. This technique was also used to encrypt information in the fractional Fourier domain [2, 3] and to store encrypted information holographically [4, 5]. Phase encoding techniques were also proposed for optical authenticity verification [6-8].

Wang *et al* [9] and Li *et al* [10] proposed another method for optical encryption and authenticity verification. Unlike the techniques mentioned above, this method encrypts information completely into a phase mask, which is located in either the input or the Fourier domain of a 4-f correlator. For instance, given the predefinitions of a significant image $f(x, y)$ as the desired output and a phase-distribution $\exp\{jb(u, v)\}$ in the Fourier domain, it's easy to optimize the other phase function $\exp\{jp(x, y)\}$ with a modified projection onto constraint sets (POCS) algorithm [10]. Therefore the image $f(x, y)$ is encoded successfully into $\exp\{jp(x, y)\}$ with the aid of $\exp\{jb(u, v)\}$. In other words, the fixed phase $\exp\{jb(u, v)\}$ serves as the lock while the retrieved phase $\exp\{jp(x, y)\}$ serves as the key of the security system. To reconstruct the original information, the phase functions $\exp\{jp(x, y)\}$ and $\exp\{jb(u, v)\}$ must match and be located in the input and the Fourier plane respectively. Abookasis *et al* [11] implemented this scheme with a joint transform correlator for optical verification.

However, because the key $\exp\{jp(x, y)\}$ contains information of the image $f(x, y)$ and the lock $\exp\{jb(u, v)\}$, and the 4-f correlator has a character of linearity, it is possible for the intruder to find out the phase-distribution of the lock function by statistically analyzing the random characters of the keys if the system uses only one lock for different image. In order to increase the secure level of such system, one approach is to use different lock function for different image. Enlarging the key space is another approach to increase the secure level. It can be achieved by encrypting images in the fractional Fourier domain; as a result, the scale factors and the transform order offer additional keys [2, 3]. On the other hand, note that the phase-mask serves as the key of the system, enlarging the key space can be achieved by encoding the target image into two or more phase masks with a modified POCS algorithm. Chang *et al* [12] have proposed a multiple-phases retrieval algorithm and demonstrated that an optical security system based on it has higher level of security and higher quality for the decrypted image. However, this algorithm retrieves only one phase-distribution with a phase constraint in each iteration. As a result, the masks are not so consistent and may affect the quality of the recovered image.

In the present paper, we propose a modified POCS algorithm that adjusts the distributions of both phase-masks synchronously in each iteration. As a result, the convergent speed of the iteration process is expected to significantly increase. And the target image with much higher quality is expected to recover because of the co-adjusting of the two masks during the iteration process. When the iteration process is finished, the target image is encoded into the phase-masks successfully. Each of these masks severs as the key of the security system and part of the encrypted image itself as well. Moreover, the algorithm can be extended to generate multiple phase-masks for arbitrary stages correlator. To acquire the maximum security, each key is assigned to different authority so that the decryption cannot be performed but being authorized by all of them. This key-assignment scheme is especially useful for military and government applications.

The algorithm description is presented in Section 2. Computer simulation of this algorithm and the corresponding discuss are presented in Section 3.

## 2. Cascaded Iterative Fourier Transform (CIFT) Algorithm

Consider the operation of the encryption system with the help of a 4-f correlator as shown in Fig.1, the phase masks placed in the input and the Fourier planes are denoted as $\boldsymbol{f}_{opt}(x,y)$ and $\boldsymbol{y}_{opt}(u,v)$, respectively, where (x, y) and (u, v) represent the space and the frequency coordinate, respectively. Once the system is illuminated with a monochromatic plane wave, a target image $f(x, y)$ (an image to be decrypted or verified) is expected to obtain at the output plane. The phase-masks $\boldsymbol{f}_{opt}(x, y)$ and $\boldsymbol{y}_{opt}(u,v)$ contain the information of $f(x, y)$, that is, $f(x, y)$ is encoded into these phase-masks. The encoding process is the optimization of the two phase-distributions. It is somewhat similar with the problems of the image reconstruction and the phase retrieval, which can be solved with the POCS algorithm. However, the present problem comes down to the phase retrieval in three (or more, in general) planes along the propagation direction. So the conventional POCS algorithm should be modified for this application.

The cascaded iteration Fourier transform (CIFT) algorithm begins with the initialization of the phase-distributions of the masks. Suppose the iteration process reaches the $k^{\text{th}}$ iteration ($k$ = 1, 2, 3, …), and the phase-distributions in the input and the Fourier plane are represented as $\boldsymbol{f}_k(x, y)$ and $\boldsymbol{y}_k(u,v)$, respectively. Then an estimation of the target image is obtained at the output of the correlator defined by

$$f_k(x, y) \exp[j\boldsymbol{j}_k(x, y)] = \text{IFT}\{\text{FT}\{\exp[j2\boldsymbol{pf}_k(x, y)]\}\exp[j2\boldsymbol{py}_k(u, v)]\}, \qquad (1)$$

where FT and IFT denote the Fourier transform and the inverse Fourier transform, respectively,. If $f_k(x, y)$ satisfies the convergent criterion, the iteration process stops, and $\boldsymbol{f}_{opt}(x, y) = \boldsymbol{f}_k(x, y)$ and $\boldsymbol{y}_{opt}(u,v) = \boldsymbol{y}_k(u,v)$ are the optimized distributions. Otherwise, the $f_k(x, y)$ is modified to satisfy the target image constraint as follows

$$f_k^{'}(x, y) = \begin{cases} f(x, y), & \text{if} \quad f(x, y) > 0 \\ f_k(x, y), & \text{if} \quad f(x, y) = 0 \end{cases}, \qquad (2)$$

Then the modified function is transformed backward to generate both of the phase-distributions as follows

$$\boldsymbol{y}_{k+1}(u,v) = \text{ang}\left\{\frac{\text{FT}\{f_k'(x,y)\exp[j\boldsymbol{j}_k(x,y)]\}}{\text{FT}\{\exp[j2\boldsymbol{p}f_k(x,y)]\}}\right\}, \quad (3a)$$

$$\boldsymbol{f}_{k+1}(x,y) = \text{ang}\left\{\text{IFT}\left\{\frac{\text{FT}\{f_k'(x,y)\exp[j\boldsymbol{j}_k(x,y)]\}}{\exp[j2\boldsymbol{p}\boldsymbol{y}_{k+1}(u,v)]}\right\}\right\}, \quad (3b)$$

where ang{ · } denotes the phase extraction function. Then $k$ is replaced by $k+1$ for the next iteration. It is shown in Eqs. 3(a) and 3(b) that both of the phase-distributions are modified in every iteration, accorded to the estimation of the target image in the present iteration. It ensures the algorithm converges with much faster speed and more consistent for the phase-masks.

In general, the convergent criterion can be the MSE or the correlation coefficient between the iterated and the target image, which are defined by

$$MSE(k) = \frac{1}{M \times N}\sum_{m=1}^{M}\sum_{n=1}^{N}\left[|f(m,n)|^2 - |f_k(m,n)|^2\right]^2, \quad (4a)$$

or

$$R(k) = \frac{\sum_{m=1}^{M}\sum_{n=1}^{N}\{f(m,n) - \text{E}[f]\}\{f_k(m,n) - \text{E}[f_k]\}}{\left\{\left[\sum_{m=1}^{M}\sum_{n=1}^{N}[f(m,n) - \text{E}[f]]^2\right]\left[\sum_{m=1}^{M}\sum_{n=1}^{N}[f_k(m,n) - \text{E}[f_k]]^2\right]\right\}^{1/2}}, \quad (4b)$$

where M×N is the size of the image, and E[ · ] denotes the mean of the image. The convergent behavior of this algorithm is similar to that of the conventional POCS. That is, the MSE reduces rapidly in the foremost few iterations, then it keeps reducing slowly till it reaches the minimum. Correspondingly, the correlation coefficient is expected to increase rapidly at first and keep increasing slowly till the stopping criterion is satisfied.

In decryption, the determined phase-masks $\boldsymbol{f}_{opt}(x,y)$ and $\boldsymbol{y}_{opt}(u,v)$ (the keys or essentially, the encrypted images) are placed in the input and the Fourier plane, respectively, and then transformed into the output plane through the correlation defined by Eq. (1). The modulus of the output is the decrypted image. The CIFT algorithm retains the property of the conventional iteration algorithm, that is, the final phase-distributions of the masks are determined by the initializations of them. Therefore different initializations will result in different distributions of $\boldsymbol{f}_{opt}(x,y)$ and $\boldsymbol{y}_{opt}(u,v)$. The target image cannot be decrypted if the keys mismatch (that is, the keys were generated from the different iteration process).

In practical system, the phases of the masks are quantized to finite levels, which might reduce the solution space and introduce noise to the recovered image. To compensate the loss of the quality, the target image can be encoded into more phase-masks to provide additional freedom for solutions searching, which means to encrypt the image with a multi-stages (cascaded) correlator. From the point of view of security, this strategy significantly enlarges the key space (because more keys were generated), and makes the intrusion more difficult. Generally, the *t*-stages correlation is defined as

$$f'(x,y) = \begin{aligned} &\text{IFT}\{\text{FT}\{\cdots\text{IFT}\{\text{FT}\{\exp[j2\boldsymbol{p}\boldsymbol{f}^{(1)}(x,y)]\}\exp[j2\boldsymbol{p}\boldsymbol{y}^{(2)}(u,v)]\}\cdots \\ &\times \exp[j2\boldsymbol{p}\boldsymbol{f}^{(t-1)}(x,y)]\}\exp[j2\boldsymbol{p}\boldsymbol{y}^{(t)}(u,v)]\} \end{aligned}, \quad (5a)$$

for *t* is even, or

$$f'(x,y) = \begin{aligned} &\text{IFT}\{\text{FT}\{\cdots\text{IFT}\{\text{FT}\{I(x,y)\}\exp[j2\boldsymbol{p}\boldsymbol{y}^{(1)}(u,v)]\}\cdots \\ &\times \exp[j2\boldsymbol{p}\boldsymbol{f}^{(t-1)}(x,y)]\}\exp[j2\boldsymbol{p}\boldsymbol{y}^{(t)}(u,v)]\} \end{aligned}, \quad (5b)$$

for *t* is odd, where the matrix *I*(x, y) represents the input plane wave, and the superscript *i* (*i*=1, 2, …, *t*) denotes the serial number of the masks in the system. The phase-distributions of these masks may be deduced by analogous analysis for Eq. 3.

## 3. Computer simulation

In this section we numerically demonstrate our general concept. A jet plane image of the size $128 \times 128$ with 256 grayscale is used as the target image as shown in Fig. 2. The sizes of both phase-masks are same as the target image. And we suppose the optical system is illuminated by a plane wave with the amplitude equating to 1. The algorithm starts with the random initialization of the two phase-masks. Then the phase functions are transformed forward and backward alternatively through the correlation defined by Eqs. (1)-(3). The algorithm converges very fast. The correlation coefficient reaches 0.99 after about 3 iterations, then it keeps increasing slowly and finally reaches 1 within 20 iterations, correspondingly, the intensity distribution of the retrieved image is extremely close to that of the target image. Rigorously, the correlation coefficient does converge but not equate to 1 no matter how many iterations the algorithm runs because no analytic solutions for Eq. (1) can be found. Here we say it REACHES 1 just because the difference between the two images is beyond the limitation of the representational precision of the digital computer. Actually, the CIFT algorithm retains the error-reducing property of the conventional POCS algorithm. The MSE keeps reducing till the local (but not global) minimum is reached. One interesting character of the CIFT algorithm is that arbitrary initializations can generate recovered images with almost same quality, and result in different distributions for the masks, as shown in Fig. 3. Therefore the optimized phase-masks can be used as the keys of the security system. Only two phase-masks, which match each other and are located in the appropriate planes of the 4-f architecture, respectively, can recover the target image. Otherwise, the output is meaningless. On the other hand, the keys $\exp[j2\pi f_{opt}(x,y)]$ and $\exp[j2\pi y_{opt}(u,v)]$, are phase-only functions, and have random-like distributions as well. These characters may introduce a high level of security because they offer a property of anti-counterfeiting. Another secure advantage of the CIFT algorithm arises in the application of authenticity verification. Instead of detecting a single correlation peak, the verification system based on the CIFT algorithm detects a significant output to determine whether or not to verify the input. So it is impossible to cause a false verification by directly illuminating the output plane bypassing the correlator because the intruder cannot generate the same pattern at the output without the knowledge of the correct phase-distributions. This is especially useful in the applications where high security is necessary. For the sake of security, the two masks are expected assigning to two persons, respectively. Therefore, the verification can be performed only under the authorizations of them both. If higher security is required, more phase-masks can be retrieved and assigned to more authorities so as to diminish the risk of being stolen of the keys.

To compare with previous methods, the CIFT algorithm and the previous methods are investigated under the same initial conditions. Let Algorithm A, B, C and D denote the methods presented in Refs. 9, 10, 12 and the algorithm proposed in the present paper, respectively. Algorithm A and B merely modify the distribution of single mask, which is located in the Fourier plane [9] or the input plane [10], respectively, while C and D employ a searching strategy of modifying the distributions of both masks during the iteration process. Figs. 3(a)-3(d) show the corresponding recovered images of these four algorithms at the $100^{th}$ iteration. It is shown that Figs. 3(c) and 3(d) have much higher quality than Figs. 3(a) and 3(b). There may be two reasons that result in this fact. First, the solution space for C and D are significantly enlarged, and consequently, it's possible to find a better solution. Second, the latter two algorithms modify the phase distributions of both masks according to the retrieved image at the present iteration. This strategy ensures better solutions and much faster convergence. However, there is still a little difference between C and D. Algorithm C alternatively modifies the phase distributions at the input and the Fourier planes. That is, it retrieves the first phase-mask at certain iteration while fixing the others, and then retrieves the second one at the next iteration, and then again it retrieves the first one. This cycle keeps going on and on till the algorithm converges. But it's not this case for the proposed algorithm. It synchronously modifies both the phase-distributions in each iteration. This change would result in faster convergence and higher recovered quality. The MSE and the correlation coefficient between the target image and the iterated images by the four algorithms, respectively, defined by Eq. (4a) and (4b), respectively, at the $100^{th}$ iteration are shown in Table 1. Apparently, The results are consistent with those we figure out from Fig. 3. We have to point out that, although both of the correlation coefficients between Fig. 3(c) and Fig. 2, and Fig. 3(d) and Fig. 2, respectively, SEEM to equate to 1, the corresponding MSEs are quite different. Table 1 shows that the quality of the iterated image obtained by the proposed CIFT algorithm is significantly better than those obtained by others methods, owing to the most effective searching strategy.

To evaluate the convergence, we set the convergent criterion R=0.998. All these algorithms are tested with an Intel® Pentium-IV 1.6 GHz PC and under the same initial condition. In this simulation, it takes 133.6720 seconds and 622 iterations for the Algorithm A, 25.0620 seconds and 100 iterations for Algorithm B and 3.9210 seconds and 15 iterations for Algorithm C, respectively, to drive R to this threshold. Comparably, it takes only 2.3590 seconds and 9 iterations for the CIFT algorithm to reach the same threshold. Apparently, as we comment in the previous paragraph, C and D have much faster convergent speed. The MSEs of the corresponding algorithm are as follows: $9.8359\times10^{-4}$ for Algorithm A, $1.8\times10^{-3}$ for Algorithm B, $1.2\times10^{-3}$ for algorithm C, and $6.4599\times10^{-4}$ for the CIFT algorithm, respectively. However, if the iteration process is kept proceeding till the MSE never decreases any more, this value is quite different for different algorithm. Table 2 shows the comparison among these algorithms when the corresponding minimum is reached. This demonstrates the highest performance of the CIFT algorithm.

## 4. Conclusion

A cascaded iterative Fourier transform (CIFT) algorithm is proposed to design phase masks for optical security applications in this paper. Compared with previous methods, the proposed CIFT algorithm adjusts the phase-distributions of both masks synchronously as well as enlarging the searching space, and consequently, has faster convergent speed and better quality for the recovered image. The proposed algorithm encodes the target image into two phase-masks, which serve as the keys of the security system as well. These keys are assigned to different persons to obtain high level of security. Only under the authorizations of all these authorities, the decrypted image can be obtained at the output.

## Acknowledgement

This work is supported by the National Natural Science Foundation of China under grant 60277027.

List of the Figures

Fig. 1. 4-f architecture of the optical security system.

Fig. 2. The target image for the computer simulation.

Fig. 3. Corresponding iterated image at the output plane at the $100^{th}$ iterations for (a) Algorithm A, (b) Algorithm B, (c) Algorithm C, and (d) the proposed CIFT algorithm.

Fig. 4. Table of all cross correlation between functions $f(x, y)$ and $y(u, v)$. Only the pairs that were designed together at the same iteration process can recover the target image.

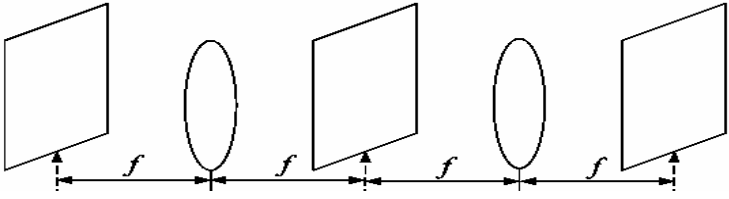

Fig. 1. 4-f architecture of the optical security system.

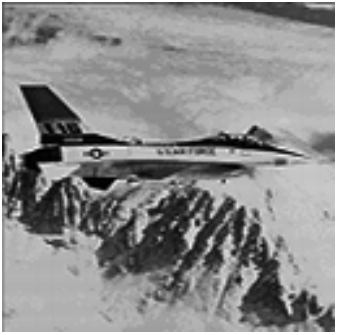

Fig. 2. The target image for the computer simulation.

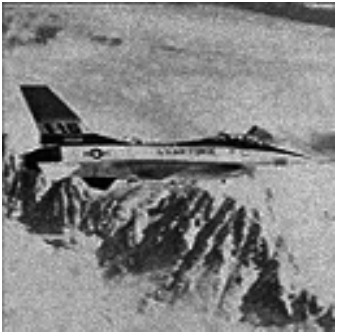 (a)

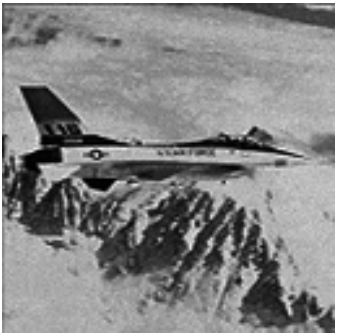 (b)

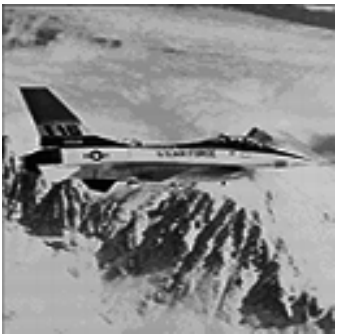 (c)

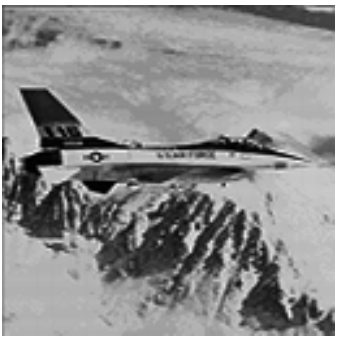 (d)

Fig. 3. Corresponding iterated image at the output plane at the 100$^{th}$ iterations for (a) Algorithm A, (b) Algorithm B, (c) Algorithm C, and (d) the proposed CIFT algorithm.

| | $\boldsymbol{y}_1$ | $\boldsymbol{y}_2$ | $\boldsymbol{y}_3$ |
|---|---|---|---|
| $\boldsymbol{f}_1$ | 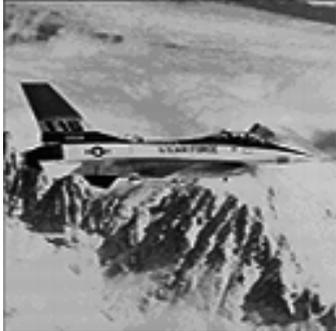 | 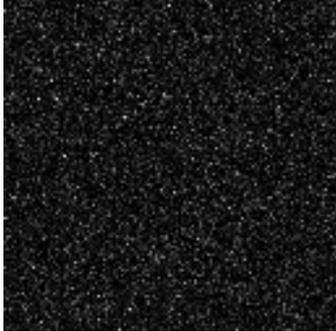 | 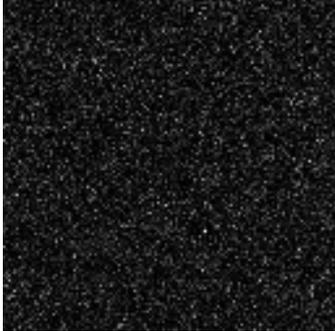 |
| $\boldsymbol{f}_2$ | 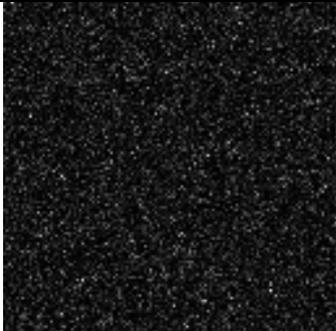 | 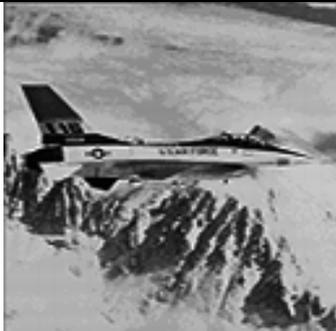 | 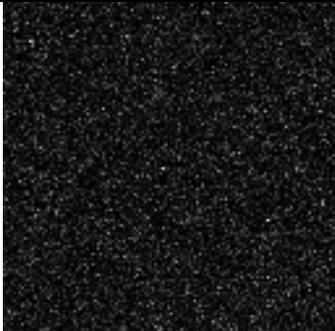 |
| $\boldsymbol{f}_3$ | 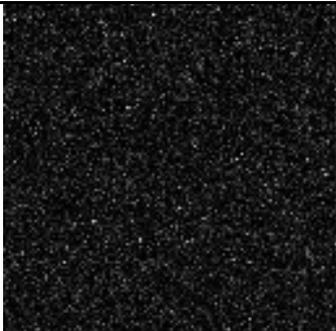 | 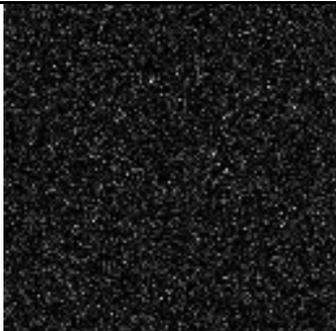 | 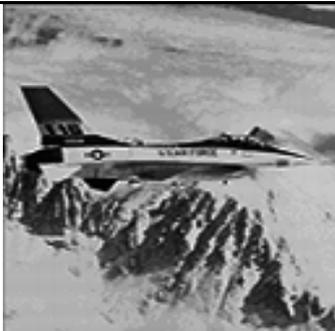 |

Fig. 4. Table of all cross correlation between functions $\boldsymbol{f}(x, y)$ and $\boldsymbol{y}(u, v)$. Only the pairs that were designed together at the same iteration process can recover the target image.

List of the tables

Table 1. The MSE and the correlation coefficient comparisons among the four algorithms at the 100$^{th}$ iteration.

| Algorithm | A | B | C | CIFT |
|---|---|---|---|---|
| MSE | 0.0064 | 0.0018 | $4.3630 \times 10^{-11}$ | $1.0896 \times 10^{-18}$ |
| R | 0.9870 | 0.9960 | 1.0 | 1.0 |

Table 2. The MSE comparison among the four algorithms when the corresponding algorithm completely converges.

| Algorithm | A | B | C | CIFT |
|---|---|---|---|---|
| MSE | $10^{-4}$ | $10^{-6}$ | $10^{-12}$ | $10^{-30}$ |